# Do Chinese Internet Users Exist Heterogeneity in Search Behavior?

Ren-jie Han[1] Shi-yuan Liu[2] Qian Li[2*]

[1]School of Economics, Chongqing Technology and Business University, Chongqing, China
[2]School of Economics, Sichuan University, Chengdu, China
***Correspondence**: School of Economics, Sichuan University, Chengdu, China
E-mail: 215582192@qq.com

Abstract: Investor attention is an important concept in behavioral finance. Many articles have conducted cross-disciplinary research leading by this concept. In this paper, we use data extraction technology to collect a large number of Baidu Index keyword search volume data[1]. After analyzing the data, we draw a conclusion that has not been paid attention to in all the past research. We find heterogeneity in searching by internet users in China. Firstly, in terms of search behavior, internet users are more inclined to use the PC end to obtain information when facing areas which need to be taken seriously by them. Secondly, attention is heterogeneous while searching. When Internet users search for information in mobile end, their attention is divergent, and search for seemingly unrelated keywords at the same time which limits their attention to information.

1. Introduction

Investor attention can't be obtained directly through observation in the behavioral finance field. With the rapid development of Internet technology and the advancement of information storage technology, massive amounts of Internet data are produced, generated and stored by Internet participants. These data, which leave traces of Internet users' access to information by means of data storage technology, can provide wide exploration space for scientific researchers.

Classical financial theory holds that asset price is the discounted value of expected return of assets. Investor attention or investor sentiment plays no role in the formation of stock prices. Rational investors will optimize the statistical characteristics of their portfolios under competitive circumstances, which will lead to a equilibrium between the price and the discount value of the expected cash flow. Meanwhile, even irrational investors do exist, whose demand will be compensated by arbitrageurs. So the price will not be affected. The capital asset pricing model assumes that information is immediately reflected in the price after it is received by market participants. The premise of this hypothesis is that as market participants, they should continue to pay full attention to their assets. However, market anomalies, such as "media effect", "earnings announcement effect" and "overconfidence", were founded in many studies, can't be explained by classical theory. In fact, attention is an extremely scarce resource. Financial market participants' attention is also a scarce resource, who have limited acceptance of information and attention.

Google and Baidu search engines, as the main tools for Internet users to access information, provide Google Trends and Baidu Index respectively. While engraving the traces of Internet users' searches, they also provide an effective way for researchers to measure investors' attention, which has become a hot area of academic research in recent years. Baidu had more than half of China's search market share, before Google withdraw from China (Oct, 2010) [1]. Since that, the share of Baidu in China's search market has remained over 70%. Through search engines, micro-individuals conduct hundreds of millions of searches every day. Internet users' collective wisdom and financial market participants can be regarded as a complex system consisting of many interactive subunits that can respond quickly to external changes [2]. Naturally, we will ask whether there is a link between the search volume of keywords in search engines and financial market volatility.

---

[1] Readers can get the code and data in this paper from us by e-mail.

The remainder of this paper is structured as follows: Section 2 reviews the representative articles by using Google Trends or Baidu index to predict financial market volatility; Section 3 is data description and discussion; Section 4 demonstrates the results; The conclusions and prospects for future research are also presented in the Section 4.

2. Literature Review

During the past decades, many studies have focused on the Baidu index and Google Trends predicting financial market volatility. In Table 2.1 lists representational researches.

**Table:2.1 Researches on Financial Market Fluctuation Based on Baidu Index (Google Trends)**

| Literature | Index Select | | Time span | Keywords | Objectives |
|---|---|---|---|---|---|
| | Baidu Index | Google Trends | | | |
| Song et al.(2011)[3] | | √ | 2005-2011 | Companies' name before IPO | the searching volume's, Companies' name before IPO(weekly data), impacts on first-day abnormal return and long term performance |
| Yu&Zhang(2012)[4] | √ | | 2011-2012 | Sum of stock abbreviations and stock code searched in Baidu index | Baidu Index Daily Search Volume Data and GEM Stock Market Performance |
| Zhang et al.(2014)[5] | √ | | 2009-2011 | stock abbreviations | Exploring the relationship between stock returns and liquidity with investors' attention which measured by Baidu search volume |
| T.Preis&Reith.D(2010)[2] | | √ | 2004-2010 | abbreviation name of listed companies in S&P500 | Through Google Trends (Weekly data) to study its relationship with stock price volatility of S&P 500 listed companies |
| Zhi Da at al.(2011)[6] | | √ | 2004-2008 | ticker of the Russell 3000 Index stock, and the IPO stock uses the company name | Relationship between Google Trends (weekly search volume) and Russell 3000 Index stocks (including IPO stocks) volatility |
| Zhang et al.(2013)[7] | √ | | 2011-2012 | Listed company name | Select 30 stocks from Shanghai Stock Exchange and Shenzhen Stock Exchange randomly and use Baidu Index to forecast price fluctuation. |
| T.Preis et al.(2013)[8] | | √ | 2004-2011 | 98 keywords related to financial markets | Using 98 keywords , studied the relationship between Google Trends and Dow Jones Index daily closing price volatility. |
| Zhou et al.(2018)[9] | √ | | 2006-2017 | 28 keywords related to financial markets and household consumption | Forecasting CSI300 volatility using the search volume daily data |

In much earlier studies, there were no direct variables that measure investors' attention. Thus, some indirect variables, such as trading volume [10][11], news and headlines [12][13][14], advertising costs [15][16], were used as investors' attention. These assumptions, which regard when the trading volume of a stock is abnormal or the code (or name) of a stock appears in the news media, investors will pay attention to the corresponding market changes, are generally very harsh. As media coverage itself, doesn't mean that investors will pay attention to. The relevant information will be worthy only when it is read by the investors. Although a large amount of information is now produced on the Internet, it is inappropriate to be an agent variable of investors' concern. Thus, it is not precise by using manufactured internet information to predict financial market volatility. Baidu Index and Google Trends represent the attention of internet users on political, economy and social life, which reflect spontaneous behavior. Besides financial market volatility predicting, Baidu Index and Google Trends have also played an important role in research of cultural communication [17], geography [18], influenza prediction [19], which provides a brand new visual angle for researchers and draws many novel conclusions.

From Table 2.1, we can find some problems by using Baidu Index and Google Trends in financial market volatility predicting.

Firstly, researchers from China mainly choose Baidu Index, while western researchers only use Google Trend. According to the annual monitoring report of Chinese search engine issued by Erie Consulting over the years, Baidu has been in a monopoly position in the Chinese search engine market. Similarly, for internet users worldwide, Google has always been the most frequently used search engine [20]. Through the Table 2.1, we can clearly figure that, the research articles that combine Baidu Index and Google Trends are rare. This could be a new direction for researchers who are interested in this field.

Secondly, Baidu Index has always been unable to be downloaded, which make it more difficult to obtain keyword data than Google Trends. Most Chinese researchers collect Baidu Index manually, which is the main reason for the short time span in their works. With the development of data extraction technology, in T. Preis et al. (2013) and Zhou et al. (2018)'s research, the number of trend (index) keywords extraction is more abundant, and the extraction span has been greatly increased than former papers.

Thirdly, recognizing search engine users' search behavior and identifying it as investors' attention for financial markets is a key point in this research field. When internet users search for "Yili (伊利)", "Mengniu (蒙牛)" or "Apple", they are more likely to purchase (or obtain services) online for these "keywords" than to collect these companies' financial market information. Since the listed company's name has multiple meanings, such as "Baiyun Airport (白云机场)", "Industrial and Commercial Bank of China(工商银行)", "Southern Airlines(南方航空)", "Amazon" or "Microsoft", which will make this situation more common. At the same time, investors may use different keywords when searching for the same companies, such as "Maotai (茅台)", "Kweichow Moutai (贵州茅台)" or 600519 (stock ticker of Kweichow Moutai ); "American Airlines", " AAR Crop", "AA" or "AAR". The inherent property of Chinese makes its information density higher than all alphabetic languages [21]. Therefore, when searching in Chinese, the collocation or synonym expression will be less than that in alphabetic languages, which is also the main difference between using Google Trend and Baidu Index to obtain keyword data. Taking the word "profit(利润)" as an example, there are five common expressions in English: profit (s), margin (s), return (s), gain (s) and earning (s), while there is only one expression in Chinese. Since tickers of listed companies' are unique, when internet users search for "BABA" or "600519", it is obvious that the main purpose of them is to obtain financial information of the corresponding companies.

Fourthly, most papers use traditional statistical methods. Y-L Zhou et al. (2018) use Long Short-Term Memory Neural Network to predict the volatility of Shanghai-Shenzhen 300 Index by regularizing 28 keywords search volume to describe the financial market. The prediction effect is better than GARCH model.

3. Data Description and Discussion

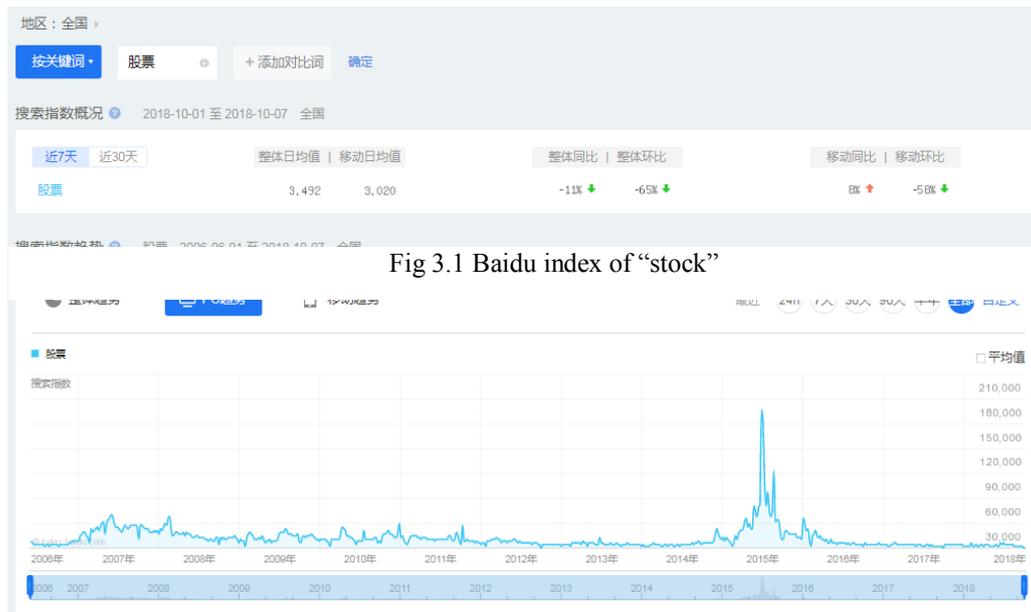

Fig 3.1 Baidu index of "stock"

Baidu Index provides daily data and weekly data of keyword search volume. When the search cycle is more than 12 months, Baidu Index returns weekly search data of keywords. Fig 3.1 is the keyword "stock" from Baidu Index. The starting time of search volume on PC end is June 1, 2006, and that on mobile end is January 1, 2011. In the left corner of Fig 3.1, there is a watermark "@index.baidu.com"[2]. For Baidu Index doesn't provide keywords downloading in any format, therefore, we complied a picture recognition program based on Python to capture the selected keyword from Baidu Index. Because of the existence of watermark, in the process of recognition, the coordinates represented by the watermark will be recognized as the search volume, resulting in 5% recognition error. For the error part, we made a manual comparison and correction for each keyword.

By referring Pries et al. (2013), and Da et al. (2015) [22], and the conclusion that Chinese semantics information in Pellegrino (2011) is higher than that in native language, this paper selects the Baidu index of 28 keywords related to macro-economy and household consumption, as shown in Table 3.1.

**Table 3.1 Keyword Selection**

| Field | Exact search Terms | Abbreviation |
|---|---|---|
| Macroeconomic keywords | financial revenue | finre |
|  | financial investment | finin |
|  | economics | econo |
|  | inflation | infla |

---

[2] When this paper is finished, the watermark and coordinate positions have changed.

|  |  |  |
|---|---|---|
|  | profit | profi |
|  | insurance | insur |
|  | bond | bond |
| Resident Consumption | travel | trave |
|  | auto buyers | autbu |
|  | luxury goods | luxgo |
|  | air ticket | airti |
|  | marriage | marri |
|  | credit card | creca |
|  | education | educa |
|  | loan | loan |
| Financial market | stock | stock |
|  | bank | bank |
|  | Financial derivatives | finde |
|  | increase | incre |
|  | advertisement | adver |
| Large consumable | real estate | reaes |
|  | office building | offbu |
|  | auto finance | autfi |
| Negative keywords | anti-corruption | antco |
|  | leverage | lever |
|  | crisis | crisi |
|  | debt | debt |
|  | default | defau |

By the end of December 2012, the number of mobile phone users in China was 420 million, accounting for 74.5% of the total number of Internet users[3]. By the end of 2017, 97.5% of Internet users use mobile phones to access the Internet, with a population of 753 million. The usage of desktop computers, laptops and tablets were all declined, and mobile phones continue to occupy the use of other personal internet devices[4]. In terms of data size, the search volume of keywords on PC end ranges from June 1, 2006 to October 29, 2017, and on mobile end from January 1, 2011 to October 29, 2017.

3.1 Descriptive Statistics

Table 3.2 and 3.3 are descriptive statistics of keywords search volume on PC end and mobile end respectively.

Table3.2 Descriptive Statistics of Keywords Searched on PC End

| Variable | Obs | Mean | Std. Dev. | Min | Max |
|---|---|---|---|---|---|
| insur | 4169 | 1790.335 | 934.1771 | 83 | 11030 |
| finre | 4169 | 156.6814 | 54.40296 | 60 | 668 |
| loan | 4169 | 1939.062 | 1030.054 | 300 | 11453 |
| antco | 4169 | 375.214 | 319.332 | 64 | 2692 |

---

[3] Data source: China Internet Information Center,《China Mobile Internet Development Statistics Report 2012》.

[4] Data source: China Internet Information Center 41st China Internet Development Statistics Report.

| Variable | Obs | Mean | Std. Dev. | Min | Max |
|---|---|---|---|---|---|
| reaes | 4169 | 2037.044 | 826.6523 | 459 | 10601 |
| debt | 4169 | 73.73759 | 101.2969 | 72 | 589 |
| lever | 4169 | 306.1667 | 185.5755 | 59 | 1983 |
| stock | 4169 | 15939.06 | 2951.57 | 1033 | 526884 |
| adver | 4169 | 2479.573 | 3722.412 | 386 | 47589 |
| airti | 4169 | 11483.03 | 6708.205 | 1925 | 51158 |
| educa | 4169 | 1363.662 | 822.2374 | 209 | 9440 |
| marri | 4169 | 1337.267 | 647.0152 | 383 | 7066 |
| finin | 4169 | 193.8067 | 143.858 | 61 | 1189 |
| finde | 4169 | 113.3293 | 79.26823 | 61 | 425 |
| econo | 4169 | 827.9826 | 209.829 | 243 | 2175 |
| profi | 4169 | 266.2115 | 121.0154 | 64 | 950 |
| trave | 4169 | 846.2219 | 391.7249 | 310 | 10252 |
| autbu | 4169 | 103.0919 | 44.69551 | 59 | 322 |
| autfi | 4169 | 154.6243 | 105.3845 | 58 | 605 |
| luxgo | 4169 | 1507.518 | 4018.955 | 130 | 52616 |
| infla | 4169 | 868.6978 | 513.1861 | 110 | 11030 |
| crisi | 4169 | 194.7255 | 78.34109 | 65 | 892 |
| defau | 4169 | 95.22121 | 46.64571 | 57 | 2702 |
| offbu | 4169 | 496.1479 | 193.6149 | 80 | 1085 |
| creca | 4169 | 3504.221 | 2478.43 | 356 | 17605 |
| bank | 4169 | 1696.533 | 563.9791 | 482 | 6086 |
| incre | 4169 | 132.5182 | 65.27357 | 60 | 836 |
| bond | 4169 | 286.4608 | 366.2367 | 176 | 2799 |

Table3.3 Descriptive Statistics of Keywords Searched on Mobile End

| Variable | Obs | Mean | Std. Dev. | Min | Max |
|---|---|---|---|---|---|
| insur | 2494 | 2226.659 | 1161.403 | 428 | 6177 |
| finre | 2494 | 181.48 | 142.3346 | 63 | 2554 |
| loan | 2494 | 5370.515 | 3293.463 | 476 | 40259 |
| antco | 2494 | 832.5926 | 987.6459 | 61 | 34762 |
| reaes | 2494 | 1420.756 | 702.0253 | 330 | 5513 |
| debt | 2494 | 235.6347 | 87.75586 | 70 | 918 |
| lever | 2494 | 327.8873 | 236.724 | 60 | 1810 |
| stock | 2494 | 14180.85 | 2084.99 | 1783 | 237727 |
| adver | 2494 | 982.1251 | 296.5515 | 63 | 2774 |
| airti | 2494 | 13565.16 | 6400.146 | 1731 | 35307 |
| educa | 2494 | 905.7285 | 170.5844 | 348 | 2388 |
| marri | 2494 | 1765.723 | 793.8147 | 849 | 6221 |
| finin | 2494 | 446.5232 | 456.1529 | 59 | 7750 |
| finde | 2494 | 171.2674 | 84.46816 | 58 | 935 |
| econo | 2494 | 1087.131 | 325.3378 | 572 | 2165 |
| profi | 2494 | 387.8994 | 170.9874 | 69 | 1038 |
| trave | 2494 | 958.087 | 537.0433 | 59 | 13968 |
| autbu | 2494 | 135.8203 | 44.02715 | 58 | 334 |
| autfi | 2494 | 136.6073 | 60.27537 | 58 | 412 |
| luxgo | 2494 | 904.3665 | 268.1083 | 166 | 4953 |
| infla | 2494 | 1237.385 | 471.2632 | 308 | 6079 |
| crisi | 2494 | 195.0702 | 97.20072 | 60 | 1254 |
| defau | 2494 | 119.8507 | 56.69388 | 57 | 570 |
| offbu | 2494 | 400.5437 | 664.8027 | 76 | 28720 |
| creca | 2494 | 7677.491 | 3688.813 | 452 | 28668 |
| bank | 2494 | 1734.799 | 592.6874 | 582 | 5305 |
| incre | 2494 | 153.5301 | 69.49692 | 63 | 689 |
| bond | 2494 | 593.3011 | 368.7324 | 69 | 1917 |

Through tables 3.2 and 3.3, the number of observations at PC end and mobile end is 4169 and 2494, due to the characteristics of data provided by Baidu Index. By comparing the mean data, we can find that the average number of keyword searches on PC end is significantly higher than that on mobile end: real estate, stock, advertising, luxury goods and office buildings. When internet users in China, want to know or consume on the field which need to spend a lot of money and require their cautious strategy, such as real estate, office buildings, automobiles, they tend to search on PC end to collect information and intelligence. This result is quite different from the findings in footnote 4.

To further validate the above conclusion, descriptive statistics are made on the search volume of PC end at the same time as mobile-side, as shown in Table 3.4.

Table3.4 Simultaneous Descriptive Statistics

| Variable | Obs | Mean | Std. Dev. | Min | Max |
|---|---|---|---|---|---|
| insur | 2494 | 1796.997 | 1110.101 | 83 | 9525 |
| finre | 2494 | 174.5802 | 52.46042 | 63 | 668 |
| loan | 2494 | 2453.838 | 949.7357 | 437 | 11453 |
| antco | 2494 | 522.7983 | 338.3907 | 99 | 2692 |
| reaes | 2494 | 1739.591 | 639.2262 | 459 | 4969 |
| debt | 2494 | 225.1367 | 44.01666 | 72 | 589 |
| lever | 2494 | 398.1163 | 179.8144 | 95 | 1983 |
| stock | 2494 | 15519.16 | 3725.96 | 1033 | 526884 |
| adver | 2494 | 1474.204 | 514.0969 | 386 | 3565 |
| airti | 2494 | 14048.25 | 7313.944 | 1925 | 51158 |
| educa | 2494 | 981.9643 | 233.1047 | 209 | 2130 |
| marri | 2494 | 1439.937 | 779.6567 | 383 | 7066 |
| finin | 2494 | 249.1728 | 132.5312 | 61 | 1189 |
| finde | 2494 | 159.99 | 40.97052 | 61 | 425 |
| econo | 2494 | 746.7394 | 153.8217 | 243 | 1336 |
| profi | 2494 | 320.5064 | 117.9069 | 76 | 950 |
| trave | 2494 | 1042.797 | 314.7876 | 468 | 10252 |
| autbu | 2494 | 114.6283 | 38.62681 | 59 | 322 |
| autfi | 2494 | 216.65 | 83.26625 | 72 | 605 |
| luxgo | 2494 | 2107.175 | 5106.368 | 228 | 52616 |
| infla | 2494 | 906.2045 | 359.0534 | 200 | 7187 |
| crisi | 2494 | 212.3244 | 75.90047 | 87 | 892 |
| defau | 2494 | 100.0549 | 55.90661 | 0 | 2702 |
| offbu | 2494 | 566.9266 | 183.9863 | 91 | 1085 |
| creca | 2494 | 4700.939 | 2501.312 | 900 | 17605 |
| bank | 2494 | 1846.781 | 609.8385 | 482 | 6086 |
| incre | 2494 | 130.7923 | 32.7862 | 66 | 354 |
| bond | 2494 | 664.9643 | 196.6039 | 176 | 2799 |

By comparing Table 3.3 and Table 3.4, it can be found that the search volume on PC end of the keyword "auto buyers" is lower than the average search volume on mobile end. The keywords' search volume on PC end, such as "real estate", "stock", "advertisement", "luxury goods" and "office building", is still higher than mobile end. Respectively, keyword "stock", the average search volume of PC and mobile end are 15519.16 and 14180.85. This result, while demonstrates that the number of mobile internet users is huge, which also confirms that internet users are more inclined to use PC end to collect data and information when they face risky financial products and large consumer goods in order to make prudent decisions.

By comparing the above three tables, we can draw a conclusion that there is heterogeneity in search behavior between mobile end and PC end internet users in China. Internet users will be more prefer to use PC for knowledge acquisition and information collection in areas involving financial markets, financial products, real estate, automobile, which need their "serious" and "prudent" treatment. Meanwhile, keywords related to macro-economy, consumer goods and negative words are more likely to be searched through mobile end by using fragmentation time. This conclusion is contrary to the rapid development of mobile devices, and also to the intuitive cognition.

In addition, the correlation analysis is reported in Table 3.5 and Table 3.6.

Table 3.5 Relevance analysis of PC end keywords (1)

|       | insur     | finre     | loan      | antco     | reaes     | debt      | lever     |
|-------|-----------|-----------|-----------|-----------|-----------|-----------|-----------|
| insur | 1         |           |           |           |           |           |           |
| finre | 0.257***  | 1         |           |           |           |           |           |
| loan  | 0.179***  | 0.576***  | 1         |           |           |           |           |
| antco | 0.085***  | 0.172***  | 0.361***  | 1         |           |           |           |
| reaes | 0.315***  | 0.052***  | -0.008    | -0.384*** | 1         |           |           |
| debt  | -0.121*** | 0.110***  | 0.379***  | 0.737***  | -0.409*** | 1         |           |
| lever | 0.081***  | 0.358***  | 0.616***  | 0.477***  | -0.198*** | 0.579***  | 1         |
| stock | 0.059***  | 0.077***  | 0.194***  | 0.122***  | 0.119***  | 0.134***  | 0.439***  |
| adver | 0.070***  | 0.086***  | -0.028*   | -0.189*** | 0.157***  | -0.233*** | -0.050*** |
| airti | 0.100***  | 0.484***  | 0.458***  | 0.228***  | -0.062*** | -0.005    | 0.301***  |
| educa | 0.198***  | -0.235*** | -0.343*** | -0.351*** | 0.578***  | -0.377*** | -0.360*** |
| marri | 0.178***  | 0.412***  | 0.353***  | -0.188*** | 0.102***  | -0.292*** | 0.045***  |
| finin | 0.062***  | 0.222***  | 0.474***  | 0.467***  | -0.221*** | 0.607***  | 0.556***  |
| finde | 0.082***  | 0.583***  | 0.659***  | 0.360***  | -0.214*** | 0.330***  | 0.562***  |
| econo | 0.276***  | 0.101***  | 0.061***  | -0.392*** | 0.584***  | -0.462*** | -0.181*** |
| profi | 0.609***  | 0.553***  | 0.547***  | 0.462***  | -0.042*** | 0.285***  | 0.477***  |
| trave | -0.021    | 0.321***  | 0.542***  | 0.413***  | -0.293*** | 0.414***  | 0.486***  |
| autbu | 0.042***  | 0.344***  | 0.418***  | 0.161***  | -0.176*** | 0.069***  | 0.381***  |
| autfi | 0.095***  | 0.415***  | 0.641***  | 0.611***  | -0.352*** | 0.750***  | 0.767***  |
| luxgo | 0.114***  | 0.163***  | 0.177***  | -0.057*** | 0.063***  | -0.101*** | 0.084***  |
| infla | 0.408***  | 0.368***  | 0.242***  | 0.115***  | 0.111***  | -0.007    | 0.198***  |
| crisi | 0.110***  | 0.400***  | 0.447***  | 0.002     | -0.079*** | -0.018    | 0.294***  |
| defau | 0.133***  | 0.244***  | 0.218***  | 0.078***  | 0.066***  | 0.063***  | 0.177***  |
| offbu | 0.262***  | 0.462***  | 0.686***  | 0.331***  | 0.187***  | 0.413***  | 0.606***  |
| creca | 0.512***  | 0.365***  | 0.473***  | 0.670***  | -0.255*** | 0.451***  | 0.477***  |
| bank  | 0.282***  | 0.579***  | 0.548***  | 0.056***  | 0.236***  | -0.090*** | 0.234***  |
| incre | 0.140***  | 0.084***  | 0.051***  | 0.002     | 0.180***  | 0.040***  | 0.076***  |
| bond  | -0.151*** | 0.174***  | 0.400***  | 0.684***  | -0.421*** | 0.853***  | 0.619***  |

Table 3.5 Relevance analysis of PC end keywords (2)

| | stock | adver | airti | educa | marri | finin | finde |
|---|---|---|---|---|---|---|---|
| stock | 1 | | | | | | |
| adver | 0.041*** | 1 | | | | | |
| airti | 0.279*** | -0.01 | 1 | | | | |
| educa | 0.141*** | 0.056*** | -0.251*** | 1 | | | |
| marri | -0.051*** | 0.153*** | 0.342*** | -0.156*** | 1 | | |
| finin | 0.221*** | -0.129*** | 0.083*** | -0.358*** | 0.060*** | 1 | |
| finde | 0.001 | 0.103*** | 0.545*** | -0.582*** | 0.394*** | 0.315*** | 1 |
| econo | 0.120*** | 0.148*** | -0.027* | 0.446*** | 0.137*** | -0.131*** | -0.281*** |
| profi | 0.035** | 0.019 | 0.431*** | -0.250*** | 0.244*** | 0.311*** | 0.628*** |
| trave | 0.02 | -0.053*** | 0.404*** | -0.516*** | 0.136*** | 0.332*** | 0.573*** |
| autbu | 0.401*** | 0.076*** | 0.557*** | -0.370*** | 0.302*** | 0.300*** | 0.414*** |
| autfi | 0.181*** | -0.122*** | 0.285*** | -0.552*** | 0.040*** | 0.673*** | 0.627*** |
| luxgo | 0.003 | -0.012 | 0.177*** | -0.051*** | 0.277*** | 0.01 | 0.219*** |
| infla | 0.040*** | 0.239*** | 0.124*** | -0.151*** | 0.218*** | 0.151*** | 0.351*** |
| crisi | 0.078*** | 0.039** | 0.318*** | -0.308*** | 0.519*** | 0.214*** | 0.385*** |
| defau | 0.082*** | 0.034** | 0.113*** | -0.001 | 0.125*** | 0.126*** | 0.197*** |
| offbu | 0.302*** | -0.036** | 0.304*** | -0.090*** | 0.239*** | 0.560*** | 0.448*** |
| creca | 0.052*** | -0.096*** | 0.397*** | -0.383*** | 0.065*** | 0.380*** | 0.533*** |
| bank | 0.180*** | 0.094*** | 0.656*** | 0.053*** | 0.415*** | 0.008 | 0.497*** |
| incre | 0.149*** | -0.003 | -0.036** | 0.315*** | 0.118*** | 0.284*** | -0.093*** |
| bond | 0.140*** | -0.242*** | 0.152*** | -0.400*** | -0.127*** | 0.607*** | 0.396*** |

Table 3.5 Relevance analysis of PC end keywords (3)

| | econo | profi | trave | autbu | autfi | luxgo | infla |
|---|---|---|---|---|---|---|---|
| econo | 1 | | | | | | |
| profi | -0.062*** | 1 | | | | | |
| trave | -0.216*** | 0.390*** | 1 | | | | |
| autbu | 0.041*** | 0.226*** | 0.229*** | 1 | | | |
| autfi | -0.303*** | 0.565*** | 0.606*** | 0.337*** | 1 | | |
| luxgo | 0.054*** | 0.185*** | 0.099*** | 0.163*** | 0.063** | 1 | |
| infla | 0.125*** | 0.460*** | 0.091*** | 0.267*** | 0.213*** | 0.088*** | 1 |
| crisi | 0.174*** | 0.283*** | 0.305*** | 0.428*** | 0.328*** | 0.295** | 0.199*** |
| defau | 0.130*** | 0.236*** | 0.101*** | 0.129*** | 0.159*** | 0.110*** | 0.157*** |
| offbu | 0.209*** | 0.501*** | 0.429*** | 0.284*** | 0.637*** | 0.134** | 0.197*** |
| creca | -0.226*** | 0.834*** | 0.450*** | 0.279*** | 0.603*** | 0.110*** | 0.381*** |
| bank | 0.290*** | 0.499*** | 0.281*** | 0.357*** | .159*** | 0.252*** | 0.242*** |
| incre | 0.193*** | 0.079*** | -0.142*** | -0.017 | -0.007 | 0.002 | 0.025 |
| bond | -0.420*** | 0.332*** | 0.482*** | 0.111*** | 0.760*** | -0.100*** | -0.009 |

Table 3.5 Relevance analysis of PC end keywords (4)

| | crisi | defau | offbu | creca | bank | incre | bond |
|---|---|---|---|---|---|---|---|
| crisi | 1 | | | | | | |
| defau | 0.160*** | 1 | | | | | |
| offbu | 0.337*** | 0.251*** | 1 | | | | |

|       | creca    | bank     | incre    | bond     |         |          |   |
|-------|----------|----------|----------|----------|---------|----------|---|
| creca | 0.216*** | 0.125*** | 0.375*** | 1        |         |          |   |
| bank  | 0.368*** | 0.253*** | 0.465*** | 0.321*** | 1       |          |   |
| incre | 0.008    | 0.131*** | 0.267*** | -0.045***| 0.178***| 1        |   |
| bond  | 0.082*** | 0.091*** | 0.496*** | 0.458*** | 0.008   | 0.109*** | 1 |

Table 3.6 Relevance analysis of Mobile end keywords (1)

|       | insur    | finre    | loan     | antco    | reaes    | debt     | lever    |
|-------|----------|----------|----------|----------|----------|----------|----------|
| insur | 1        |          |          |          |          |          |          |
| finre | 0.244*** | 1        |          |          |          |          |          |
| loan  | 0.447*** | 0.321*** | 1        |          |          |          |          |
| antco | 0.038*   | 0.069*** | 0.281*** | 1        |          |          |          |
| reaes | 0.592*** | 0.177**  | 0.714*** | 0.275*** | 1        |          |          |
| debt  | 0.544*** | 0.299*** | 0.655*** | 0.153*** | 0.602*** | 1        |          |
| lever | 0.469*** | 0.248*** | 0.709*** | 0.188*** | 0.640*** | 0.649*** | 1        |
| stock | 0.121*** | -0.003   | 0.291*** | 0.186*** | 0.210*** | 0.220*** | 0.563*** |
| adver | 0.566*** | 0.137*** | 0.544*** | 0.121*** | 0.655*** | 0.600*** | 0.623*** |
| airti | 0.040**  | 0.112*** | 0.528*** | 0.455*** | 0.418*** | 0.272*** | 0.341*** |
| educa | 0.457*** | 0.207*** | 0.548*** | 0.148*** | 0.496*** | 0.636*** | 0.682*** |
| marri | 0.311*** | 0.226*** | 0.676*** | 0.249*** | 0.557*** | 0.425*** | 0.592*** |
| finin | 0.366*** | 0.127*** | 0.555*** | 0.150*** | 0.507*** | 0.479*** | 0.596*** |
| finde | 0.275*** | 0.537*** | 0.489*** | 0.062*** | 0.212*** | 0.420*** | 0.276*** |
| econo | 0.200*** | 0.531*** | 0.329*** | 0.006    | 0.057*** | 0.225*** | 0.118*** |
| profi | 0.757*** | 0.374*** | 0.245*** | 0.071*** | 0.258*** | 0.463*** | 0.280*** |
| trave | 0.283*** | 0.077*** | 0.425**  | 0.180*** | 0.347*** | 0.351*** | 0.379*** |
| autbu | 0.349*** | 0.158*** | 0.714*** | 0.359*** | 0.704*** | 0.501*** | 0.633*** |
| autfi | 0.522*** | 0.272*** | 0.825*** | 0.245*** | 0.738*** | 0.761*** | 0.763*** |
| luxgo | 0.420*** | 0.145*** | 0.568*** | 0.217*** | 0.528*** | 0.457*** | 0.450*** |
| infla | 0.666**  | 0.294*** | 0.060*** | -0.050** | 0.118*** | 0.310*** | 0.146*** |
| crisi | 0.231*** | 0.204*** | 0.513*** | 0.183*** | 0.379*** | 0.422*** | 0.593**  |
| defau | 0.200*** | 0.495*** | 0.494*** | 0.075*** | 0.196*** | 0.377*** | 0.263*** |
| offbu | 0.175*** | 0.082*** | 0.226*** | 0.046**  | 0.178*** | 0.158*** | 0.269**  |
| creca | 0.717*** | 0.307*** | 0.698*** | 0.301*** | 0.642*** | 0.682*** | 0.647*** |
| bank  | 0.369*** | 0.212*** | 0.828*** | 0.401*** | 0.703*** | 0.650*** | 0.728*** |
| incre | 0.185*** | 0.201*** | 0.500*** | 0.215*** | 0.351*** | 0.401*** | 0.570*** |
| bond  | 0.430*** | 0.263*** | 0.745*** | 0.239*** | 0.580*** | 0.670*** | 0.825*** |

Table 3.6 Relevance analysis of Mobile end keywords (2)

|       | stock     | adver    | airti    | educa    | marri    | finin    | finde    |
|-------|-----------|----------|----------|----------|----------|----------|----------|
| stock | 1         |          |          |          |          |          |          |
| adver | 0.135***  | 1        |          |          |          |          |          |
| airti | 0.307***  | 0.333*** | 1        |          |          |          |          |
| educa | 0.398***  | 0.537*** | 0.255*** | 1        |          |          |          |
| marri | 0.251***  | 0.385*** | 0.417*** | 0.423*** | 1        |          |          |
| finin | 0.293***  | 0.420*** | 0.210*** | 0.425*** | 0.482*** | 1        |          |
| finde | 0.029     | 0.242*** | 0.203*** | 0.322*** | 0.287*** | 0.141*** | 1        |
| econo | -0.066*** | 0.091*** | 0.104*** | 0.170*** | 0.229*** | 0.040**  | 0.894*** |
| profi | 0.062*** | 0.348*** | 0.070*** | 0.391*** | 0.128*** | 0.180*** | 0.439*** |
| trave | 0.185*** | 0.316*** | 0.369*** | 0.320*** | 0.370*** | 0.353*** | 0.148*** |
| autbu | 0.287*** | 0.625*** | 0.673*** | 0.423*** | 0.653*** | 0.525*** | 0.203*** |
| autfi | 0.269*** | 0.661*** | 0.397*** | 0.593*** | 0.663*** | 0.629*** | 0.361*** |
| luxgo | 0.119*** | 0.645*** | 0.647*** | 0.416*** | 0.357*** | 0.284*** | 0.273*** |
| infla | -0.007 | 0.202*** | -0.181*** | 0.266*** | 0.047** | 0.073*** | 0.294*** |
| crisi | 0.413*** | 0.301*** | 0.335*** | 0.500*** | 0.563*** | 0.351*** | 0.350*** |
| defau | -0.026 | 0.227*** | 0.237*** | 0.264*** | 0.335*** | 0.138*** | 0.843*** |
| offbu | 0.190*** | 0.142*** | 0.092*** | 0.212*** | 0.241*** | 0.163*** | 0.132*** |
| creca | 0.266*** | 0.593*** | 0.449*** | 0.588*** | 0.506*** | 0.515*** | 0.405*** |
| bank | 0.438*** | 0.546*** | 0.687*** | 0.607*** | 0.643*** | 0.557*** | 0.322*** |
| incre | 0.406*** | 0.276*** | 0.408*** | 0.497*** | 0.581*** | 0.341*** | 0.309*** |
| bond | 0.577*** | 0.474*** | 0.394*** | 0.694*** | 0.713*** | 0.562*** | 0.394*** |

Table 3.6 Relevance analysis of Mobile end keywords (3)

| | econo | profi | trave | autbu | autfi | luxgo | infla |
|---|---|---|---|---|---|---|---|
| econo | 1 | | | | | | |
| profi | 0.379*** | 1 | | | | | |
| trave | 0.077*** | 0.241*** | 1 | | | | |
| autbu | 0.079*** | 0.165*** | 0.494*** | 1 | | | |
| autfi | 0.192*** | 0.300*** | 0.421*** | 0.724*** | 1 | | |
| luxgo | 0.152*** | 0.355*** | 0.481*** | 0.606*** | 0.498*** | 1 | |
| infla | 0.285*** | 0.779*** | 0.086*** | -0.052*** | 0.138*** | 0.135*** | 1 |
| crisi | 0.236*** | 0.208*** | 0.283*** | 0.427*** | 0.516*** | 0.306*** | 0.131*** |
| defau | 0.833*** | 0.334*** | 0.150*** | 0.207*** | 0.354*** | 0.274*** | 0.200*** |
| offbu | 0.119*** | 0.085*** | 0.104*** | 0.161*** | 0.221*** | 0.120*** | 0.084*** |
| creca | 0.280*** | 0.640*** | 0.489*** | 0.613*** | 0.733*** | 0.600*** | 0.437*** |
| bank | 0.156*** | 0.227*** | 0.482*** | 0.778*** | 0.787*** | 0.590*** | 0.023 |
| incre | 0.213*** | 0.151*** | 0.317*** | 0.445*** | 0.499*** | 0.303*** | 0.051** |
| bond | 0.227*** | 0.298*** | 0.388*** | 0.598*** | 0.779*** | 0.397*** | 0.190*** |

Table 3.6 Relevance analysis of Mobile end keywords (4)

| | crisi | defau | offbu | creca | bank | incre | bond |
|---|---|---|---|---|---|---|---|
| crisi | 1 | | | | | | |
| defau | 0.339*** | 1 | | | | | |
| offbu | 0.266*** | 0.119*** | 1 | | | | |
| creca | 0.440*** | 0.363*** | 0.216*** | 1 | | | |
| bank | 0.564*** | 0.332*** | 0.213*** | 0.736*** | 1 | | |
| incre | 0.620*** | 0.323*** | 0.261*** | 0.444*** | 0.582*** | 1 | |
| bond | 0.730*** | 0.361*** | 0.323*** | 0.652*** | 0.772*** | 0.710*** | 1 |

According to the results reported in tables 3.5 and 3.6, the correlation coefficient between keywords search volume is basically less than 0.8, which indicates that there is no strong correlation between keywords both in PC end and mobile end. The correlation coefficients of "education" with "financial derivatives" and "automobile finance" in PC end are - 0.55, while in mobile end, the correlation coefficients are 0.322 and 0.593 respectively.

This shows that, for the same keywords, there are differences in search behavior and habits between different terminal internet users.

In PC end, the keywords with high correlation with "real estate" are "education" (0.578), "economy" (0.584), "bond" (-0.421); while in mobile end, the keywords with high correlation with "real estate" become "debt"(0.602), "leverage" (0.64), "advertising"(0.655), "education"(0.496), "marriage" (0.557), "auto buyers"(0.704), "auto finance"(0.738), "credit card"(0.642),"bank" (0.703), "bond"(0.58). In the mobile end, the correlation coefficient between keywords that seem unrelated is very high. This shows mobile users tend to search more diversified, which also proves their lack of attention.

3.2 Discussion

After comparing the above keywords, we can find several rules. Firstly, the correlation of mobile keyword search volume is generally higher than that of in PC end. Secondly, for example, the keyword "real estate", in the PC end, has a strong correlation with the keywords, which have certain inherent search logic between them. While in mobile end, there is also a high correlation between "real estate" and the keywords without any logical connection.

4. Conclusion

In this paper, we find heterogeneity in searching by internet users in China. Firstly, in terms of search behavior, internet users are more inclined to use the PC end to obtain information when facing areas which need to be taken seriously by them. Secondly, attention is heterogeneous while searching. When Internet users search for information in mobile end, their attention is divergent, and search for seemingly unrelated keywords at the same time which limits their attention to information. Therefore, although the number of mobile end users has far exceeded that of PC end users, because of the heterogeneity of search behavior and attention concentration in information acquisition, if the research field needs to be taken seriously by information acquirers, the Baidu Index of PC end is chosen as proxy variable, which more likely has lower data noise and more accurate results.

In order to test the conclusion of this paper, different algorithms can be introduced to apply the data of PC and mobile terminals, and the prediction accuracy of different algorithms can be compared. This is the main limitation of our paper. In future research, we will use different algorithms to validate this conclusion.